\documentclass[]{interact}

\usepackage{epstopdf}
\usepackage[caption=false]{subfig}

\usepackage[numbers,sort&compress]{natbib}
\bibpunct[, ]{[}{]}{,}{n}{,}{,}
\renewcommand\bibfont{\fontsize{10}{12}\selectfont}
\makeatletter
\def\NAT@def@citea{\def\@citea{\NAT@separator}}
\makeatother

\theoremstyle{plain}

\theoremstyle{definition}

\theoremstyle{remark}

\begin{document}


\title{Role of vortical structures for enstrophy and scalar transport in flows with and without stable stratification}

\author{
\name{M.~M. Neamtu-Halic\textsuperscript{a,b}\thanks{CONTACT M.~M. Neamtu-Halic. Email: neamtu@ifu.baug.ethz.ch}, J.-P. Mollicone\textsuperscript{c}, M. van Reeuwijk\textsuperscript{c} and M. Holzner\textsuperscript{b,d}}
\affil{\textsuperscript{a}Institute of Environmental Engineering, ETH Zürich, CH-8039 Zürich, Switzerland; \\
         \textsuperscript{b}Swiss Federal Institute of Forest, Snow and Landscape Research WSL, 8903 Birmensdorf, Switzerland;\\
         \textsuperscript{c}Department of Civil and Environmental Engineering, Imperial College London, London SW7 2AZ, UK;\\
        \textsuperscript{d}Swiss Federal Institute of Forest, Snow and Landscape Research WSL, 8903 Birmensdorf, Switzerland }
}

\maketitle

\begin{abstract}
In this paper, we investigate the enstrophy dynamics in relation to objective Eulerian coherent structures (OECSs) and their impact on the enstrophy and scalar transport near the turbulent/non-turbulent interface (TNTI) in flows with and without stable stratification. We confirm that vortex-stretching produces enstrophy inside the boundaries of the OECSs, while viscous diffusion transfers the enstrophy across the boundaries of the structures. Although often overlooked in the literature, viscous dissipation of enstrophy within the boundaries of vortical structures is significant. Conversely, for the weakly stratified flows also investigated here, the effect of the baroclinic torque is negligible. We provide evidence that the OECSs advect the passive/active scalar and redistribute it via molecular diffusion. Finally, we use conditional analysis to show that the typical profiles of the enstrophy and scalar transport equation terms across the TNTI are compatible with the presence of OECSs positioned at the edge between the turbulent sublayer and the turbulent core region. We show that when these profiles are further conditioned to the presence of OECSs, their magnitude is considerably higher.
\end{abstract}


\section{\label{sec:intro}Introduction}

Vortical coherent structures have received considerable attention from the turbulence community over the last decades. Mostly defined as regions with concentrated vorticity and material coherence with a life time larger than the typical time scales of the flow \cite{dubief2000coherent,haller2015lagrangian}, vortical structures constitute an appealing tool for understating complex turbulent flow phenomena  \citep{lesieur1987turbulence}. Moreover, vortical structures are valuable to investigate the mixing and the transport of mass, momentum and scalar (e.g. temperature or concentration) in turbulent flows \citep{haller2015lagrangian}.  

In turbulent flows, a well-established convention divides vortical structures into two classes: the so-called large scale vortices (LSVs) and the intense vorticity structures (IVSs) \citep{da2011intense}. The LSVs originate from the particular instability of a certain type of flow and their characteristics such as size and lifetime are reported to be flow dependent \citep{frisch1995turbulence, tsinober2009informal}. On the other hand, IVSs have similar characteristics across a variety of flows. Mostly detected through a threshold on the vorticity \citep{jimenez1993structure}, IVSs are known in isotropic turbulence as “worms” \citep{siggia1981numerical}. Many studies dedicated to these worms showed that universally their size is of order of $5\eta$, with $\eta$ the Kolmogorov length scale, in isotropic turbulence \citep{siggia1981numerical,jimenez1993structure,jimenez1998characteristics,vincent1991satial}, in mixing layers \citep{tanahashi2001appearance}, in channel flows \citep{kang2007dynamics} and in jets \citep{ganapathisubramani2008investigation}. In an early study by \citet{jimenez1993structure}, IVSs were observed to be rather stable structures and their dynamical behaviour was shown to be similar to that of a stable Burger vortex model, which is characterized by a radial balance inside a vortex tube between enstrophy production and enstrophy diffusion. 
However, a direct assessment of the enstrophy production and enstrophy diffusion mechanisms is still missing in the literature. Moreover, it is not clear how other mechanisms such as viscous dissipation of enstrophy or the baroclinic torque that is present in stratified flows may contribute to these. To date, research progress on the enstrophy transport by vortical structures has been hampered by arbitrariness in the detection methods, which are mostly based on thresholding either the vorticity field \citep{hussain1986coherent,jimenez1993structure,da2011intense} or the vorticity relative to the strain field \citep{okubo1970horizontal,hunt1988eddies,weiss1991dynamics,hua1998exact}. Progress made in a recent string of research \citep{haller2015lagrangian} allows to overcome the arbitrariness of the classic methods and permits to identify objective (i.e. observer-independent) coherent structures, as required for replicable experiments. 
 
In the present work, we extract objective coherent structures and investigate the enstrophy dynamics inside vortical structures to shed a light on the mechanisms that govern the time evolution of the enstrophy contained in these structures. We use a newly developed extraction method to systematically identify objective Eulerian coherent structures (OECSs) \citep{haller2016defining,serra2016objective} and we apply it to direct numerical simulations (DNSs) data of a turbulent flow with and without stable stratification.  

The impact of vortical structures on the transport of scalars, such as concentration or temperature, have been matter of numerous studies in recent years \citep{kadoch2011role,beta2003wavelet}. The main motivation has been to better understand the role of vortical structures in organizing scalar transport which is relevant in many practical applications, e.g. mixing of pollutants in atmosphere or heat-transfer in heat exchangers and gas turbines. In particular, \citet{kadoch2011role} showed that in homogeneous, isotropic turbulence vortical structures are mainly responsible for turbulent transport and mixing of passive scalars. The impact of vortical structures on heat transport was investigated by \citet{dharmarathne2018coherent} in a thermal turbulent channel flow. They observed that vortical structures near the wall contribute to the removal of hot fluid from the wall to outer region. \citet{debusschere2004turbulent} studied heat transport in plane channel and Couette flows. Their results indicated that in channel flow the overall vertical heat transfer is lower as compared to Couette flow. They attributed this observation to the presence of large scale vortical structures in Couette flow that transport heat across the center line of the flow, while similar structures are missing in plane channel flow. \citet{frohlich2008scalar} investigated the scalar transport in co-annular swirling jets. They found that $30-40\%$ of the concentration fluctuation is carried by large-scale coherent flow structures. Here, we use conditional analysis to investigate the impact of OECSs on the transport and diffusion of the active/passive scalar of flows with and without stable stratification.

At the boundaries of turbulence, a sharp and highly contorted interface, so-called turbulent/non-turbulent interface (TNTI), is known to separate the turbulent region form the irrotational surrounding fluid \citep{corrsin1955free, westerweel2009momentum, da2014interfacial}. The ambient surrounding fluid is continuously entrained into the turbulent side through the TNTI, a phenomenon known as turbulent entrainment. It has been shown that turbulent entrainment is a two-stage process \citep{chauhan2014turbulent, mistry2016entrainment, watanabe2016lagrangian}. Initially, at the outer edge of the TNTI, a non-turbulent fluid parcel acquires vorticity via viscous diffusion \citep{holzner2011laminar} and subsequently, in the turbulent region vorticity is amplified through vortex stretching. The role played by the vortical structures along these two stages was recently investigated by \citet{watanabe2017role}. The authors used a model to show that the average profile of enstrophy production and enstrophy viscous diffusion near the TNTI are compatible with the presence of Burger-type vortex near the interface. By positioning a Burger-vortex of the size of the IVSs at a distance of approximately $9\eta$ from the TNTI, they predicted reasonably well the profiles of the enstrophy production and enstrophy viscous diffusion across the TNTI of a free shear turbulent flow. However, in its present form their model lacks other effects such as viscous dissipation of enstrophy and baroclinic torque that may potentially be of significance. In this work, we compute the conditional profiles of the terms in the enstrophy transport equation across the TNTI and compare them to those crossing an OECS near the interface. The goal is to understand the impact of vortical structures as extracted from the flow on the dynamics of enstrophy near the TNTI. To identify the OECSs that are in proximity of the TNTI and to investigate their contribution to enstrophy and scalar transport, we use a recently developed conditional analysis by \citet{neamtu2019lagrangian}. 

The turbulent transport of the scalar across the TNTI is an important phenomenon for many applications of practical interest (e.g. chemical reactors) \citep{dimotakis2000mixing}. In a recent work by \citet{watanabe2015turbulent}, it was shown that a large jump of passive scalar exists at the boundary of turbulent flow regions and that molecular diffusion exchanges the passive scalar between the turbulent region and the fluid in the TNTI proximity. Since vortical structures are known to carry large amount of scalar \citep{kadoch2011role}, larger gradients of the scalar are expected in proximity of the TNTI when OECSs are present. Our aim is to understand how OECSs impact the transport of the scalar near the TNTI.

In nature, turbulent flows develop frequently in presence of stable stratification (e.g. cloud-top mixing layers, river plumes and oceanic overflows). In these flows, the entrainment rate is known to diminish with increasing Richardson number $Ri$, the ratio between the buoyancy and shear strength of the flow \citep{ellison1959turbulent}. Nowadays, it is widely accepted that entrainment rate reduces with increasing stratification as a consequence of the reduction of both the local entrainment velocity and the area of the TNTI \citep{krug2015turbulent, van2018small, van2018mixing}. Recently, \citet{neamtu2019lagrangian} used experimental data of a gravity current to show that OECSs modulate the area of the TNTI and that their modulating capacity diminish with increasing stratification. Moreover, OECSs have been observed to organize the flow field on the TNTI proximity thereby imposing the local entrainment velocity \citep{neamtu2019lagrangian} and setting the mechanism that produces/destroys the surface area of the nearby TNTI \citep{neamtu2020evolution}. It remains to be understood how OECSs affect the process of the entrainment and how do they adapt at different levels of stratification. To this end, we apply the conditional analysis of the enstrophy transport equation to the direct numerical simulations data of gravity currents and of a wall-jet. Moreover, we investigate how the diffusion of the active/passive scalar across the interface in the OECSs proximity varies with increasing stratification.

The main scope of the paper is to investigate the dynamics of the enstrophy inside the OECSs and to understand the role of the OECSs in the transport of the scalar, with a particular regard to the region in proximity of the TNTI. 

The paper is organized as follows. In \S \ref{sec:methods} we present the DNS data set. This is followed by the presentation of the results in \S\ref{sec:results}, while a summary and concluding remarks are given in \S\ref{sec:summary}.


\section{\label{sec:methods}Methods}

\subsection{\label{sec:DNS}DNS data set}

\begin{table}
\tbl{Simulation parameters: $N_{i}$ and $L_{i}$ denote the number of grid points and the size along $i$-direction respectively.  The subscript $\textit{0}$ indicates the inflow parameters. The Taylor Reynolds number $Re_{\lambda}=\sqrt{15/ \nu \epsilon} e^{1/2}$ is computed averaging the rate of turbulent dissipation $\epsilon$ and the turbulent kinetic energy $e$ over $120<t<130$.}
{\begin{tabular}{lcccccc} \toprule

  $ $  & $\alpha (deg.)$ & $Ri_{0}$ & $Re_{0}$ & $Re_{\lambda}$ & $N_{x} N_{y} N_{z}$ & $L_{x} L_{y} L_{z} / h_{0}^{3}$ \\ \midrule
       $Ri0$    & $-$ & $0$ & $3700$ & $115$ & $1536^{2} \times 1152$ & $20^{2} \times 10$\\[1ex]
       $Ri11$   & $10$ & $0.11$ & $3700$ & $105$ & $1536^{2} \times 1152$ & $20^{2} \times 10$\\[1ex]
       $Ri22$   & $5$ & $0.22$ & $3700$ & $70$ & $1536^{2} \times 1152$ & $20^{2} \times 10$\\[1ex]\bottomrule
\end{tabular}}
\label{table:tab1}
\end{table}

\begin{figure}
  \centerline{\includegraphics[width=0.9\linewidth]{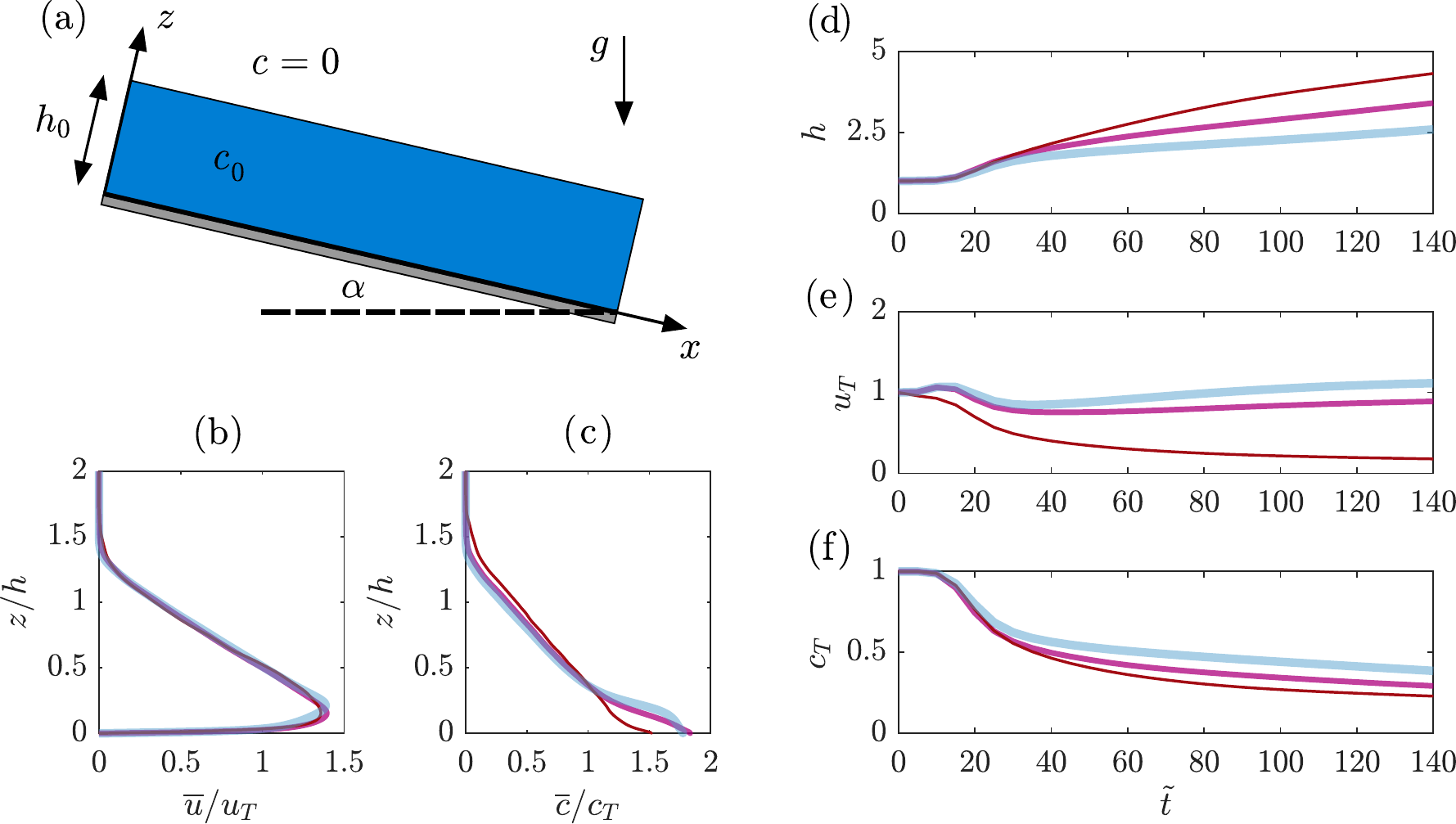}}
  \caption{Schematic representation of the simulation setup (a). Vertical profiles of the mean streamwise velocity (b) and mean concentration (c) for $Ri0$ (red), $Ri11$ (purple) and $Ri22$ (light blue). Time variation of current height (c), top-hat velocity (d) and top-hat concentration (e).}
\label{fig:fig1}
\end{figure}

The data set employed in this work consists of DNSs of temporally evolving gravity currents and of a temporal turbulent wall-jet. These flows are particularly suitable to study the transport of enstrophy and scalar in that different intensities of  vertical transport of these quantities can be obtained by varying the stratification level. It is thus possible to investigate how OECSs adapt to this change and contribute to transport.  
The simulations reproduce the classical experiment of \citet{ellison1959turbulent}, in which a lighter turbulent fluid flows along the top of an inclined wall in a heavier ambient irrotational fluid. As sketched in figure \ref{fig:fig1}(a), we reverse the problem upside-down as we simulate the motion of a negatively buoyant fluid flowing down a slope inclined at an angle $\alpha$. The physics of the problem is unaffected as we consider a Boussinesq fluid. The temporal problem is particularly suitable for obtaining converged statistics relatively inexpensively, as it is homogeneous in the wall-normal planes and the statistics depend only on time and wall normal direction. For the simulations, we employ SPARKLE, a code that solves the Navier–Stokes equations in the Boussinesq approximation \citep{craske2015energy}

\begin{equation}
  \frac{\partial \boldsymbol{u}}{\partial t} + \boldsymbol{u} \cdot \nabla \boldsymbol{u} = -\nabla p+\nu \nabla^{2} \boldsymbol{u}+\boldsymbol{b},
  \label{equation:mom} 
\end{equation}

\begin{equation}
  \frac{\partial c}{\partial t} + \boldsymbol{u} \cdot \nabla c = D\nabla^{2}c,
  \label{equation:con} 
\end{equation}

\begin{equation}
  \nabla \cdot \boldsymbol{u}=0.
  \label{equation:Sc1} 
\end{equation}

with a fourth-order accurate finite volume discretization scheme \citep{craske2015energy} on a cuboidal domain. Here, $\boldsymbol{u}=(u,v,w)$ is the fluid velocity in the $x$ streamwise, $y$ spanwise and $z$ wall-normal direction, $p$ is the (modified) kinematic pressure, $\boldsymbol{b}=\beta \boldsymbol{g} c$ is the buoyancy, with $\boldsymbol{g}=($sin$\alpha,0,$cos$\alpha)$ to simulate the sloping bottom and $\beta=\rho_{0}^{-1} \partial \rho/\partial c|_{c_{0}}$, and $\nu$, $D$ are the kinematic and molecular diffusivity, respectively. For the wall jet $c$ is a passive scalar (with Schmidt number $Sc=1$), whilst for the gravity current $c$ is an active scalar. 

The boundary conditions are periodic in the streamwise and the spanwise direction, while in the vertical direction, at the wall ($z=0$) and at the top of the simulation domain, no slip and free slip velocity boundary conditions are imposed respectively for the velocity and Neumann (no-flux) boundary conditions are imposed for scalar c. For the initial conditions (indicated by subscript 0), a uniform distribution of both the streamwise velocity $u_{0}$  and the scalar $c_{0}$ up to a height $h_{0}$ above the bottom wall are implemented. A schematic representation of the simulation set-up is shown in figure \ref{fig:fig1}(a). The size of the domain is $L_{x} \times L_{y} \times L_{z}=20h_{0} \times 20h_{0} \times 10h_{0}$ with a resolution of $1536^{2} \times 1152$. For a more detailed discussion on the DNSs concept and numerical configuration we refer to \citet{van2018small,van2018mixing}. 

Following \citet{neamtu2020evolution}, we simulated three different flow cases, namely a wall-jet ($\beta=0$) and two different gravity currents ($\beta>0$). The flow cases differ in the initial Richardson number $Ri_{0}=B_{0}$ cos $(\alpha)/u_{0}^{2}$, where $B_{0}=\beta g c_{0} h_{0}$ is a conserved quantity in the simulations, whereas the initial bulk Reynolds number $Re_{0}=u_{0}h_{0}/\nu$ is kept constant. Table \ref{table:tab1} summarizes the parameters of the simulations employed in this study. Note that the label of the flow cases indicates the value of $Ri_{0}$. The results presented here are based on data over six independent $xz$-planes, which are equally spaced in the $y$-direction, amounting to 280 snapshots over a period of 140$\tilde{t}$, with $\tilde{t}h_{0}/u_{0}$. 
 
Throughout the paper, we use the following top-hat definitions

\begin{equation}
  u_{T}h= \int_{0}^{\infty} \overline{u}dz, \,\,\,\,\,\,\,\,\,\,\,\,\,\,  u_{T}^{2} h= \int_{0}^{\infty} \overline{u}^{2}dz \,\,\,\,\,\,\,\,\,\,\,\,\,\, \textrm{and} \,\,\,\,\,\,\,\,\,\,\,\,\,\,  c_{T}h= \int_{0}^{\infty} \overline{c}dz,
\end{equation}
\\
where $\overline{u}$ and $\overline{c}$ are the mean streamwise velocity respectively the mean concentration, computed averaging in wall-parallel planes.

To characterize the structure of the flows, in figure \ref{fig:fig1}(b) the mean profile of the streamwise velocity are normalized with the top hat definitions, showing a collapse on a single curve  for all flow cases in the outer layer of the flow. That is, although there are fundamental differences between the gravity currents and the wall-jet, the structure of the flows is similar. The mean profile of the scalar concentration is shown in figure \ref{fig:fig1}(c). Again, when normalized with the top hat definitions, the profiles collapse on a single curve, although $c$ has a different physical interpretation in the stratified cases as compared to the unstratified one. We note that the similarities of the flow structure shown here are due to the fact that although the gravity currents presented here are buoyancy driven, the turbulence characteristics are shear dependent \citep{krug2017fractal}.

The time evolution of the top-hat definitions is shown in figure \ref{fig:fig1}(d-f). After an initial transition, the current height $h$ grows linearly for the gravity currents and $h\propto \tilde{t}^{1/2}$ for the wall-jet, while $u_{T}$ is constant for the gravity currents and $u_{T}\propto \tilde{t}^{-1/2}$ for $Ri_{0}$ \citep{van2018small}. As $Ri$ increases, $h$ decreases while $u_{T}$ increases. On the other hand, $c_{T}\propto \tilde{t}^{-1}$ for the gravity currents and $c_{T}\propto \tilde{t}^{-1/2}$ for the wall-jet, with $c_{T}$ increasing with decreasing stratification. 

\subsection{\label{sec:OECSsEdTNTIid}OECSs eduction and TNTI identification}

The identification method used in this work to extract vortical structures is based on the so-called instantaneous vorticity deviation (IVD). The IVD is an observer-independent scalar field that measures an intrinsic material rotation rate of fluid elements \citep{haller2016defining}. Derived from a new dynamic version of the classic polar decomposition \citep{haller2016dynamic}, the IVD field is defined by 

\begin{equation}
  \textit{IVD}( \boldsymbol{x},t) =| \boldsymbol{\omega} ( \boldsymbol{x} ,t)- \boldsymbol{\overline{\omega}} (t)| 
  \label{equation:eqIVD} 
\end{equation}

where $\omega(\boldsymbol{x},t)$ is the vorticity vector and $\overline{\omega}(,t)$ is its spatial mean. In particular, the $\it{IVD}$ provides an observer-independent local angular velocity for each point of the fluid mass and therefore, enables the identification of OECSs in an observer-independent manner, as required for reproducible coherent structure extraction \citep{haller2015lagrangian}. Local maxima of the $\it{IVD}$ field identify the center of the OECSs, while the outermost almost convex contour of the $\it{IVD}$ encircling a local maximum represents the boundary of the structure.
To contain the computational costs, we extract 2D OECSs from vertical planes of 3D data. We use a criterion defined in \citet{neamtu2020evolution} in which a maximum of the $\it{IVD}$ field is selected only if the ratio between the two eigenvalues of the Hessian of $\it{IVD}$ at the location of $\it{IVD}$ maxima is below a threshold. The rationale behind this criterion is based on the fact that a 2D OECS in a slice results form the intersection of the tubular structures with the plane itself. Since most of the dynamics of tubular vortical structures happens in planes perpendicular to the center-line of the structure, we select only structures that are perpendicular to the plane \citep{neamtu2020evolution}. The results presented throughout the paper corresponds to $120<\tilde{t}<130$.

To identify the position of the TNTI, we impose a threshold on the enstrophy field $\omega^{2}=\omega_{i}\omega_{i}$ \citep{bisset2002turbulent,holzner2007small,holzner2008lagrangian,silva2018scaling,
neamtu2019lagrangian}. Here, the threshold is selected as $\omega_{thr}^{2} = 10^{-3}\overline{\omega^{2}}$, where $\overline{\omega^{2}}$ is the space average of the enstrophy. Following a criterion by \citet{taveira2013lagrangian}, the threshold values were verified to lie within the interval of possible values in which there is no appreciable variation of the volume fraction of the turbulent region with $\omega_{thr}^{2}$.

\subsection{\label{sec:RadialProf}Radial profiles and conditional profiles.}

We investigate the radial profiles inside the OECSs of the terms of the enstrophy transport equation

\begin{equation}
  \frac{D\omega^2}{Dt}=2\omega_{i}\omega_{j}S_{ij}+\nu\nabla\cdot(\nabla\omega^2)-2\nu\nabla\omega_{i}:\nabla\omega_{i}+2\epsilon_{ijk}\omega_{i}\frac{\partial g_{k}^{\prime}}{\partial x_{j}}
  \label{equation:Ens} 
\end{equation}

where $\mathcal{P}_{\omega^{2}}=2\omega_{i}\omega_{j}S_{ij}$ is the the enstrophy production, with $S_{ij}$ the rate of strain tensor, $\mathcal{D}_{\omega^{2}}=\nu\nabla\cdot(\nabla\omega^2)$ is the viscous diffusion of enstrophy, $\mathcal{E}_{\omega^{2}}=-2\nu\nabla\omega_{i}:\nabla\omega_{i}$ is the viscous dissipation term, $\mathcal{B}_{\omega^{2}}=2\epsilon_{ijk}\omega_{i}\frac{\partial g_{k}^{\prime}}{\partial x_{j}}$ is the baroclinic torque, and of the scalar transport equation

\begin{equation}
  \frac{Dc}{Dt}=D\nabla\cdot(\nabla c).
  \label{equation:Sc} 
\end{equation}

\begin{figure}
  \centerline{\includegraphics[width=1\linewidth]{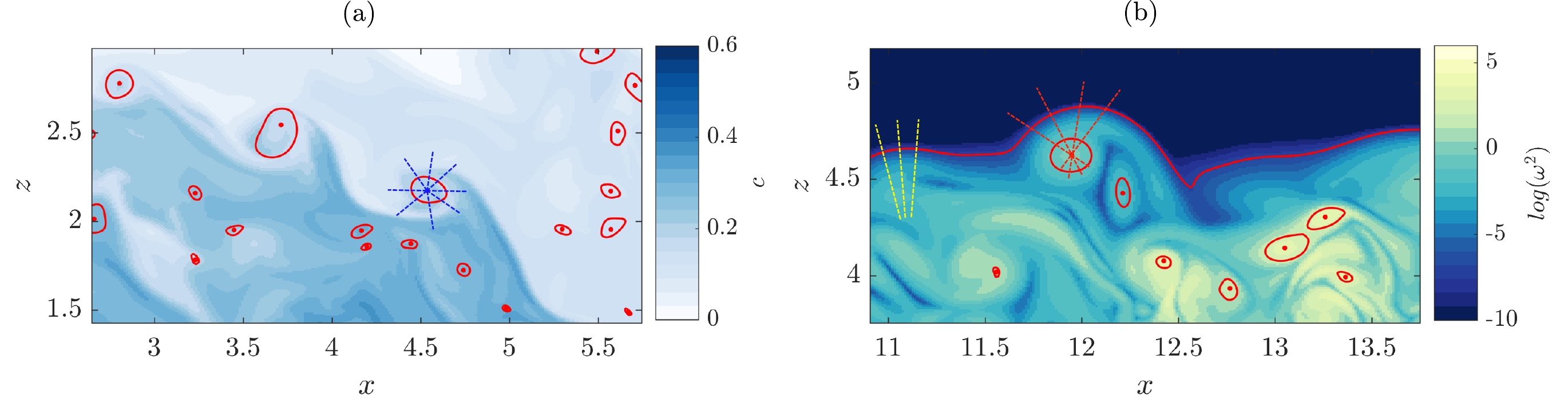}}
  \caption{(a) OECSs (red) and radial lines (dashed blu lines), with the scalar field in the background. Local normal lines (yellow) to the TNTI (red open curve) and local approximately-normal lines (red dashed lines) passing trough the center of an OECSs, with the enstrophy field in the background.}
\label{fig:fig2}
\end{figure}

In order to compute these profiles, for each selected structure we connect the center of the structure with the points on the boundaries (figure \ref{fig:fig2}a) with a segment and interpolate the scalar terms of (\ref{equation:Ens},\ref{equation:Sc}) above on the points of the segment. The distance from the center of the OECS is then normalized with the distance between the center and the boundary $R$. Note that the material derivatives are computed from the terms on the right hand side of (\ref{equation:Ens},\ref{equation:Sc}).

In the TNTI proximity, average values of the terms in equations (\ref{equation:Ens}) and (\ref{equation:Sc}) are also computed conditioned with respect to the position of the TNTI. As consolidated in the literature \citep{krug2015turbulent, da2014interfacial}, these terms are computed along lines that are normal to the TNTI itself, as schematically shown in figure \ref{fig:fig2}(b), with the distance form the interface normalized by the Kolmogorov length scale $\eta$. Here the normal to the interface points in the direction of the turbulent region. In a second set of results presented in this work, we use a further condition. That is, we compute the conditional averages of the quantities in equations (\ref{equation:Ens}) and (\ref{equation:Sc}) with respect to the TNTI along lines that are approximately perpendicular ($90^{\circ} \pm 30^{\circ}$) to the TNTI and passing through the center of OECSs near the TNTI. A sketch of this concept is illustrated in figure \ref{fig:fig2}(b). Also in this case the origin is fixed at the TNTI position.
 
\section{\label{sec:results}Results}

\subsection{\label{sec:GenDesct}General description of coherent structures}

\begin{figure}
  \centerline{\includegraphics[width=0.6\linewidth]{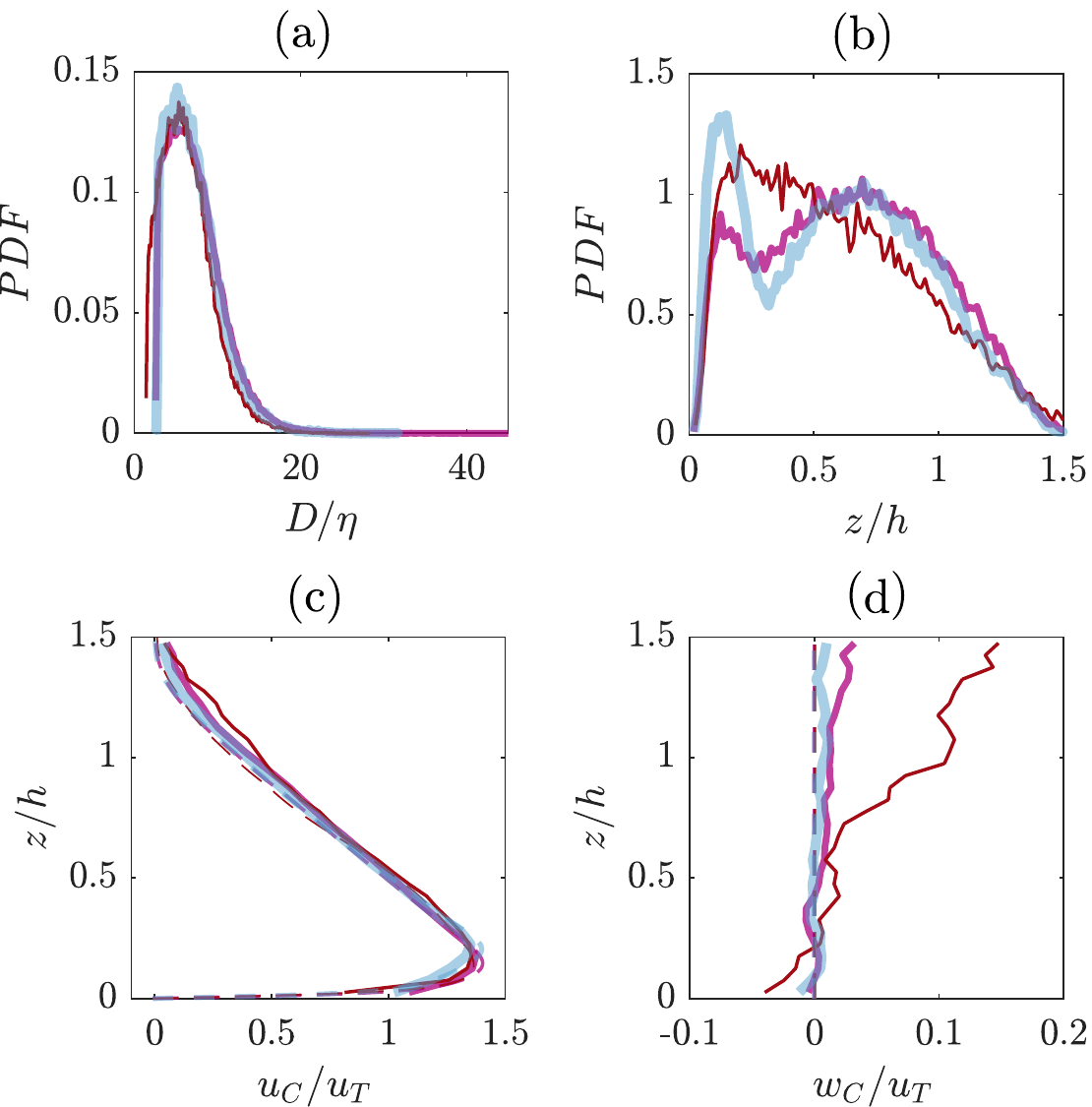}}
  \caption{PDFs of size of the OECSs (a) and of the position of their center (b). Mean streamwise (c) and wall-normal (d) velocities at the center of the OECSs (continuous) against the mean velocity profiles (dashed) for $Ri0$ (red), $Ri11$ (purple) and $Ri22$ (light blue).}
\label{fig:fig3}
\end{figure}

In the following we provide a general description of the OECSs extracted from the three flow cases. In figure \ref{fig:fig3}(a), we show the probability density function (PDF) of the size of the structures $D$, normalized by the Kolmogorov length scale $\eta=(\nu \epsilon)^{1/4}$. To compute $D$ we fit an ellipse on the boundaries of the OECSs and calculate $D$ as the mean value between the minor and the major axes of the fitting ellipse. Note that the use of a single single size to express the dimension of the OECSs is justified by the fact the aspect ratio of the structures is not far from 1, with an average value of approximately 0.8. 
The PDF of the OECSs size shows a sharp increase from approximately $3\eta$ to $7\eta$ where it presents a maximum, to decrease more gradually up to $40\eta$. As the $Ri$ number increases, there is no significant difference in terms of the size distribution of the structures. 

The position of the center of the structures normalized by the current height $h$ is shown in figure \ref{fig:fig3}(b). For all flow cases, in the wall proximity, the PDFs grow approximately linearly up to $z/h\approx0.15$, where the PDF shows a peak. Interestingly, this height corresponds to the position of the maximum of the mean stream-wise velocity profile, which means that this region is a particularly active region for the formation of coherent flow structures. From the maximum towards larger wall distance, a different behaviour can be noticed for the unstratified case compared to the stratified ones. For $Ri0$, the PDF decreases monotonically towards the outer region of the current at $z/h\approx1.5$. Conversely, for the gravity currents, the PDFs show a second peak in the center of the mixing layer region at about $z/h=0.7$. The two peaks of the PDFs of the gravity currents indicate that two different population of OECSs are present in the flow. As discussed in \citep{van2018small}, the turbulence production is zero at the velocity maximum, implying that the boundary layer becomes `decoupled' from the outer layer. This is corroborated in figure \ref{fig:fig1} by the increasing concentration difference between the boundary layer and outer layer with increasing stratification.

The mean streamwise and wall-normal velocity profiles built with the velocities at the center of the structures are shown in figure \ref{fig:fig3}(c) and (d). For a comparison, we show also the unconditioned mean velocity profiles. As shown in figure \ref{fig:fig3}(c), the structures follow the mean flow for $z/h<0.5$, while they are on average somewhat faster than the mean flow in the streamwise direction near the outer region of the current. This is consistent with the behaviour in the wall-normal direction. For $z/h>0.5$, the OECSs tend, on average, to move upwards. This means that they move from a region with a higher streamwise velocity to a region with a lower one, which makes them faster than the mean flow. Note that the upward movement of the OECSs is consistent with the growth in time of the current depth in the temporal problem. Since the structures are identified uniquely in the turbulent region, they tend to move away from the wall as the current depth grows in time. Since the vertical movement of the fluid is suppressed by stable stratification \citep{ellison1957turbulent, townsend1958turbulent}, this effect is less pronounced as $Ri$ increases. This average movement away from the wall of vortical fluid does not violate conservation of mass as non-turbulent fluid moves on average inward. That is, due to the external intermittency of these flows, the turbulent fraction moves on average outwards, while the non-turbulent fraction move inward.  Conversely, the unconditioned vertical mean velocity is zero everywhere.

\begin{figure}
  \centerline{\includegraphics[width=1\linewidth]{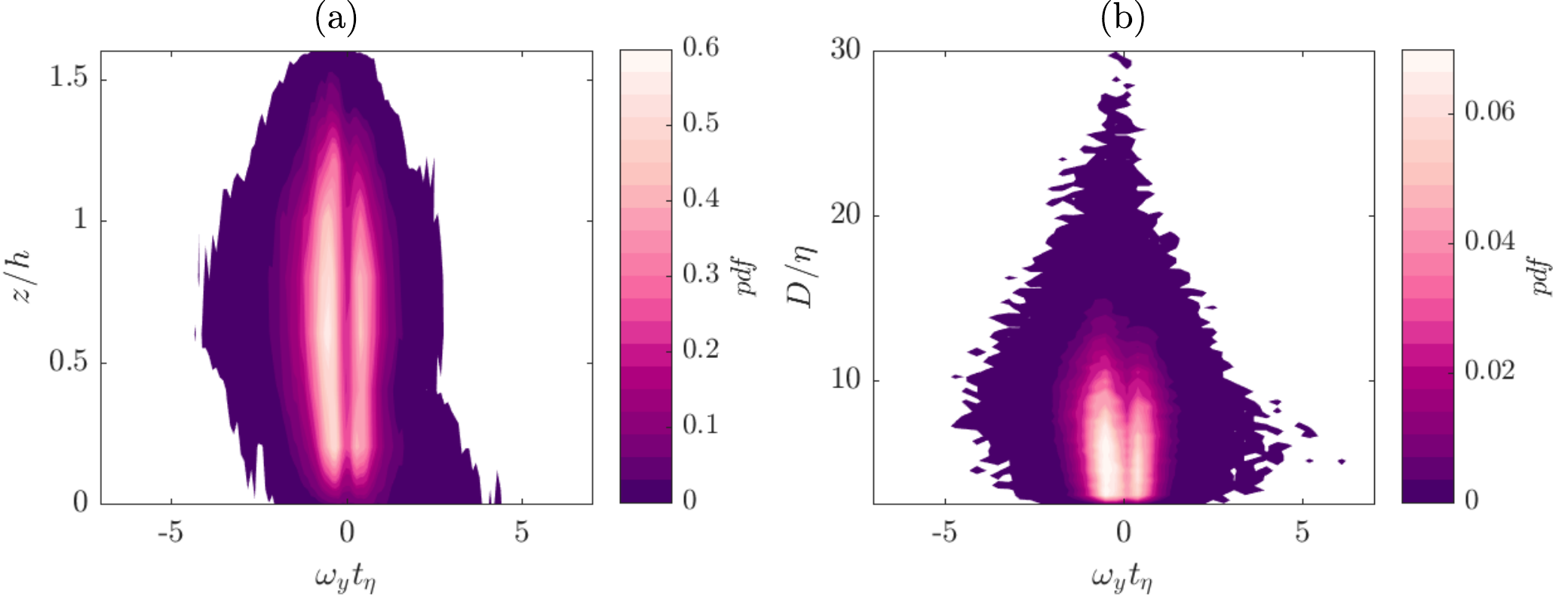}}
  \caption{JPDFs of the spanwise component of the vorticity at the center of the OECSs and the position of their center (a), respectively their size (b) for $Ri11$ flow case.}
\label{fig:fig4}
\end{figure}

In the following, we display the spanwise component of the vorticity $\omega_{y}$ at the center of the structures. We present results for $Ri11$ flow case; however qualitatively similar results can be found for the other flow cases (not shown). A joint PDF (JPDF) of $\omega_{y}$ and of the position of the center of the structures is shown in figure \ref{fig:fig4}(a). Two distinctive elongated zones of high probability can be observed. Both zones have a peak at about $z/h=0.7$, but they have opposite signs of the vorticity component. That is, the structures in this region possess either a clockwise and an anticlockwise rotations. As the outer-shape of the JPDF suggests, on average the structures tend to rotate in the counterclockwise direction (negative $\omega_{y}$) in the outer and in the mixing-layer region while they tend to rotate in the clockwise direction (positive $\omega_{y}$) in the near-wall region, which is consistent with the sign of the mean velocity gradient in these regions.

The joint PDF of $\omega_{y}$ at the center of the structures with respect to their size is shown in figure \ref{fig:fig4}(b). While the two probability peaks of $\omega_{y}$ of opposite sign are positioned at about $5\eta$, the highest vorticity is associated with structures with a size of $\approx 8-10\eta$.

\subsection{\label{sec:ImpactRadial}Radial profiles of the terms of enstrophy and scalar transport equations}

In this section, we present results for the radial profiles of the terms of the enstrophy and concentration transport equations. As explained in section \S \ref{sec:methods}, in these profiles the distance form the center of the OECSs is normalized with the distance $R$ between the center of the structures and their boundaries.

\begin{figure}
  \centerline{\includegraphics[width=1\linewidth]{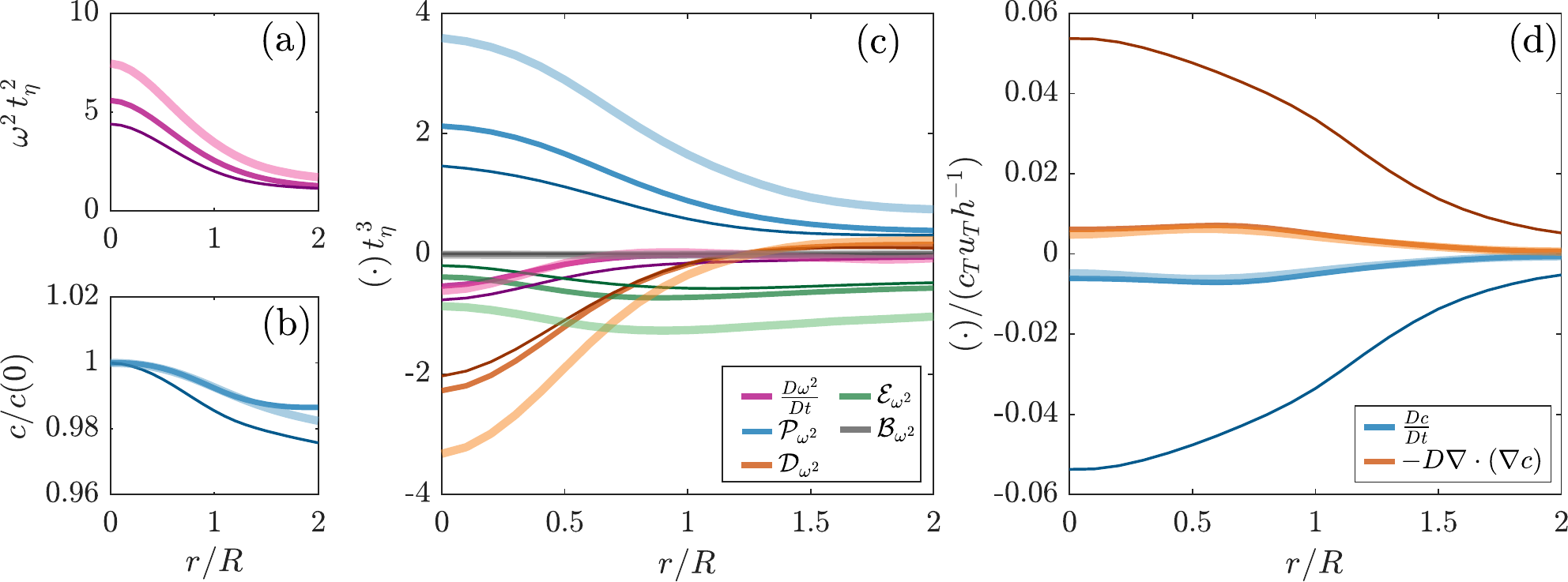}}
  \caption{Radial profiles of the enstrophy (a) and of the scalar (b) and of the terms in transport equation of enstrophy (c), respectively of the scalar (d). Increasing thickness and transparency of the lines corresponds to increasing $Ri$.}
\label{fig:fig5}
\end{figure}

In figure \ref{fig:fig5}, we show the radial profiles averaged over all the structures in the flow. Figure \ref{fig:fig5}(a) shows that the profile of the enstrophy has a bell shape with a maximum at the center of the structure, followed by a sharp decrease towards the boundary at $r/R=1$, reaching the unconditioned average value outside the boundaries of the structure at $r/R=2$. The shape of the enstrophy profile is fundamental to understand how the OECSs are sustained, in that it results from a balance of inertial and viscous effects as shown below. As $Ri$ increases, the profile flattens. This is not unexpected given the smaller magnitude of the vorticity in the whole flow with decreasing stratification. In figure \ref{fig:fig5}(b), we also show the radial profile of the scalar normalized with the value at the center of the structures $c_{0}$.  The radial profile of $c/c_{0}$ shows a similar profile as the enstrophy, although much less steep since it decreases only by a few per cent over the same span. 

In figure \ref{fig:fig5}(c), the radial profile of the terms in the enstrophy transport equation are shown. The production term $\mathcal{P}_{\omega^{2}}$ is positive with a maximum at the center of the structure and it decreases sharply towards the boundary. This is in agreement with the well-known vortex stretching mechanism, in which the enstrophy is on average produced by the interaction between vorticity $\mathbf{\omega}$ and rate of strain $S_{ij}$. Note that $\mathcal{P}_{\omega^{2}}$ has a maximum at the center of the OECSs which is consistent with the radial profile of $\omega^{2}$ shown in figure \ref{fig:fig5}(a). The viscous diffusion of the enstrophy $\mathcal{D}_{\omega^{2}}$ is negative inside the boundaries of the structure and changes sign outside the boundaries. This means that the enstrophy inside the OCESs is diffused through the boundaries of the structures to the surrounding flow. Note that the trend of $\mathcal{D}_{\omega^{2}}$ is also consistent with the radial profile of $\omega^{2}$, which shows a negative curvature for $r/R<1$ and a positive one for $r/R>1$. The viscous dissipation of the enstrophy $\mathcal{E}_{\omega^{2}}$ is negative everywhere with a minimum (in magnitude) coinciding with the center of the structures. This is again expected based on the radial profile of $\omega^{2}$, which shows little variation in the proximity of $r/R=0$, while it has a sharper decrease towards the boundaries. It is important to note that this view holds for little variation of enstrophy in the axial direction, that is for vortical structures with an elongated shape. In \citep{neamtu2019lagrangian}, we showed that 2D OECSs in vertical planes results as the intersection of tubular structures with the plane itself, justifying thereby our perspective. 

The radial profile of the baroclinc torque $\mathcal{B}_{\omega^{2}}$ shows that this term is negligible in comparison with the other terms. This means that on average, $\mathcal{B}_{\omega^{2}}$ has no relevant impact on the enstrophy transport for OECSs. The sum of all the terms, the material derivative of the enstrophy, shows a negative value inside the OECSs with a minimum at $r/R=0$, reaching an unconditioned negligible value outside the OECSs. The fact that $D\omega^{2}/Dt$ is negative inside the boundaries of the OECSs means that on average the structures decrease their enstrophy content over their evolution. In particular, it is the negative viscous diffusion of enstrophy in addition to enstrophy dissipation inside the OECSs, which is only partially balanced by the enstrophy production, that causes the decrease of $\omega^{2}$ over time. As stratification increases, all the terms of the enstrophy transport equation increase in magnitude, which is consistent with the increase of the mean enstrophy at the center of the OECSs with increasing stratification. 

In figure \ref{fig:fig5} (d), we show the radial profiles of the terms of the transport equation for the concentration. The diffusion of the concentration $D\nabla \cdot(\nabla c)$ is positive inside the structure which means that the concentration decreases in magnitude. Note that this is also consistent with the radial profile of the concentration (figure \ref{fig:fig5} b) which shows a strong positive curvature at $r/R=0.5$ that decreases towards the boundaries. Moreover, as $Ri$ increases, the diffusion is damped.  

\begin{figure}
  \centerline{\includegraphics[width=1\linewidth]{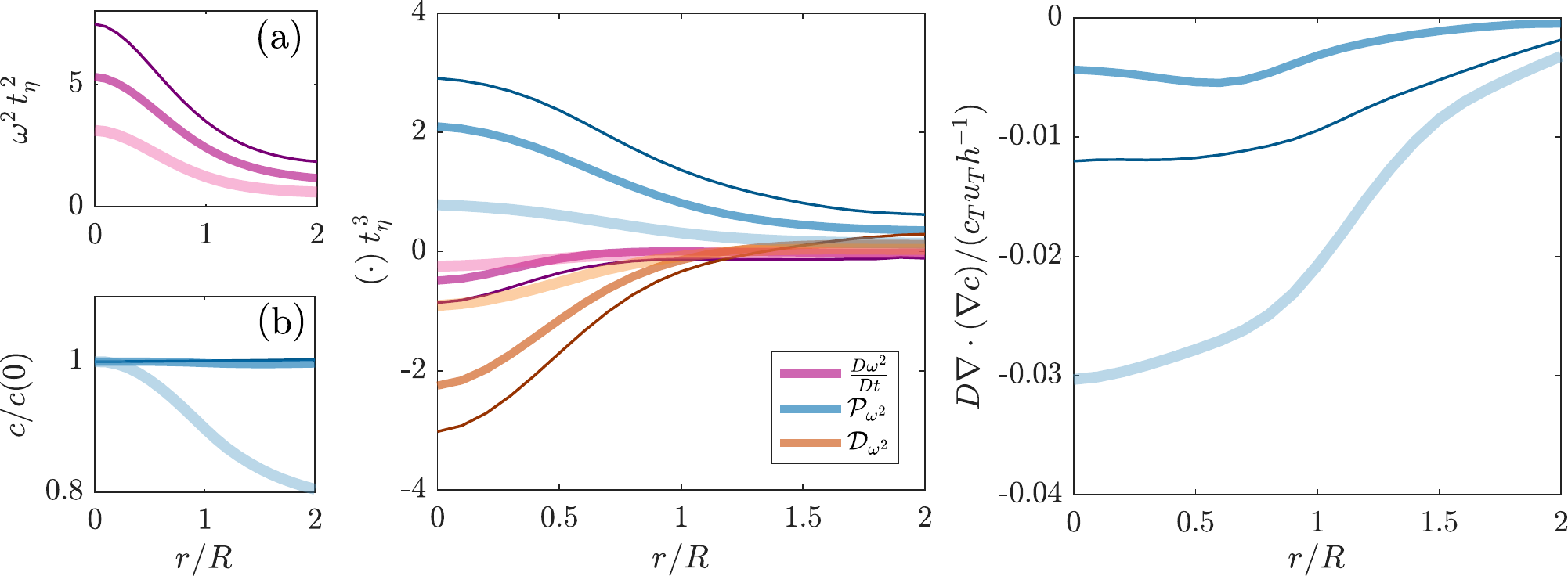}}
  \caption{Radial profiles of the enstrophy (a) and of the scalar (b) and of the terms in transport equation of enstrophy (c) and of the scalar (d) for $Ri11$ conditioned with respect to the distance of the OECSs from the wall. The thickness and the transparency of the curves increase with the distance from the wall as indicated in the text.}
\label{fig:fig6}
\end{figure}

In the following we display the same profiles shown in figure \ref{fig:fig5} conditioned with respect to the distance of the center of structures from the wall. We selected three regions: a near-wall region ($z/h<0.3$), a region where the mean shear is constant ($0.3<z/h<1.2$) and a region near the outer boundary of the current ($z/h>1.2$). In figure \ref{fig:fig6} (c), the profiles of the terms in the enstrophy transport equation show a similar behaviour to the ones presented in figure \ref{fig:fig5}(c), although with a different magnitude. They are more intense in the near-all region and decrease with increasing distance from the wall. This is in agreement with the radial profiles of enstrophy (figure \ref{fig:fig6}a), which show more intense values in the near-wall region. Indeed, as observed above, the intensity of mean velocity gradient is maximum near the wall ($z/h<0.15$), it decreases in the mixing layer ($0.3<z/h<1.2$) and it is minimum in the outer region. Differently, the radial profile of the scalar diffusion is more intense in the outer region and is minimum in the region between $0.3<z/h<1.2$ (figure \ref{fig:fig6}d). 

\begin{figure}
  \centerline{\includegraphics[width=1\linewidth]{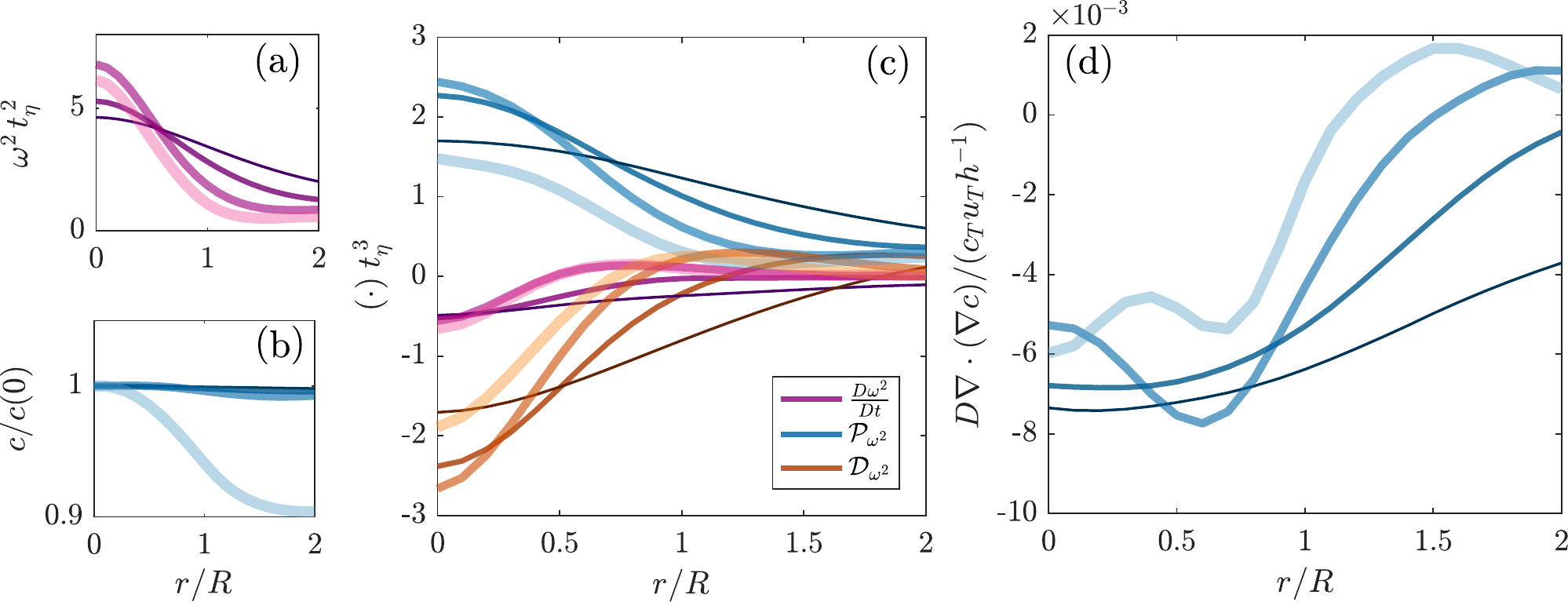}}
  \caption{Radial profiles of the enstrophy (a) and of the scalar (b) and of the terms in budget equation of enstrophy (c) and of the scalar (d) for $Ri11$ conditioned with respect to the size of the OECSs. The thickness and the transparency of the curves increase with increasing size of the structures, which are divided in the following groups: $D/\eta<4.5$, $4.5<D/\eta<8$, $8<D/\eta<12$ and $D/\eta>12$.}
\label{fig:fig7}
\end{figure}

The radial profiles of the terms of enstrophy and scalar transport equation conditioned with respect to the size of the structures are shown in figure \ref{fig:fig7}. The limits of the intervals in which the OECSs are grouped are the following: $D/\eta<4.5$, $4.5<D/\eta<8$, $8<D/\eta<12$ and $D/\eta>12$. In figure \ref{fig:fig7}(a), we show the radial profiles of the enstrophy. The enstrohy at the center of OECSs increase with the size of the structures to reach a maximum for $8<D/\eta<12$, to then decrease again for $D/\eta>12$. The average size of former interval is $10\eta$, which is the average size of the most intense vortical structures in this work and observed by others \citep{jimenez1993structure,da2011intense}. The profiles of the terms of the enstrophy transport equation follow a similar trend in which the most "active" OECSs are those in the interval $8<D/\eta<12$ (figure \ref{fig:fig7} c). These OECSs are also the more stable structures, in that they present a minimum of $Dw^2/Dt$. A different trend can be observed for the scalar diffusion. The radial profiles of the scalar transport equation show that the magnitude of the scalar molecular diffusion decreases with increasing size of the structures (figure \ref{fig:fig7} d). At first sight, this result seems to be in contradiction with the curvature of the radial profiles of the scalar (figure \ref{fig:fig7} b). However, note that while the radial profiles are normalized with $R$, the curvature of the profile it is not. That is, a match between the profile of the scalar and the magnitude of the scalar molecular diffusion cannot be expected. The results are qualitatively similar  for $Ri0$ and $Ri22$ (not shown).

\subsection{\label{sec:ImpactRadialnearTNTI}Impact of coherent structures on the enstrophy and concentration transport near the TNTI}

In the following, we investigate the impact of the OECS on the terms of the enstrophy and scalar transport terms conditioned with respect to the TNTI position. The normal distance $\tilde{z}$ from the TNTI is normalized by the Kolmogorov length scale $\eta$ as it is common in the literature \citep{da2014interfacial, krug2015turbulent, silva2018scaling}  and the normal to the interface is oriented in the turbulent direction such that $\tilde{z}/\eta$ is positive inside the turbulent region and negative outside.

\begin{figure}
  \centerline{\includegraphics[width=1\linewidth]{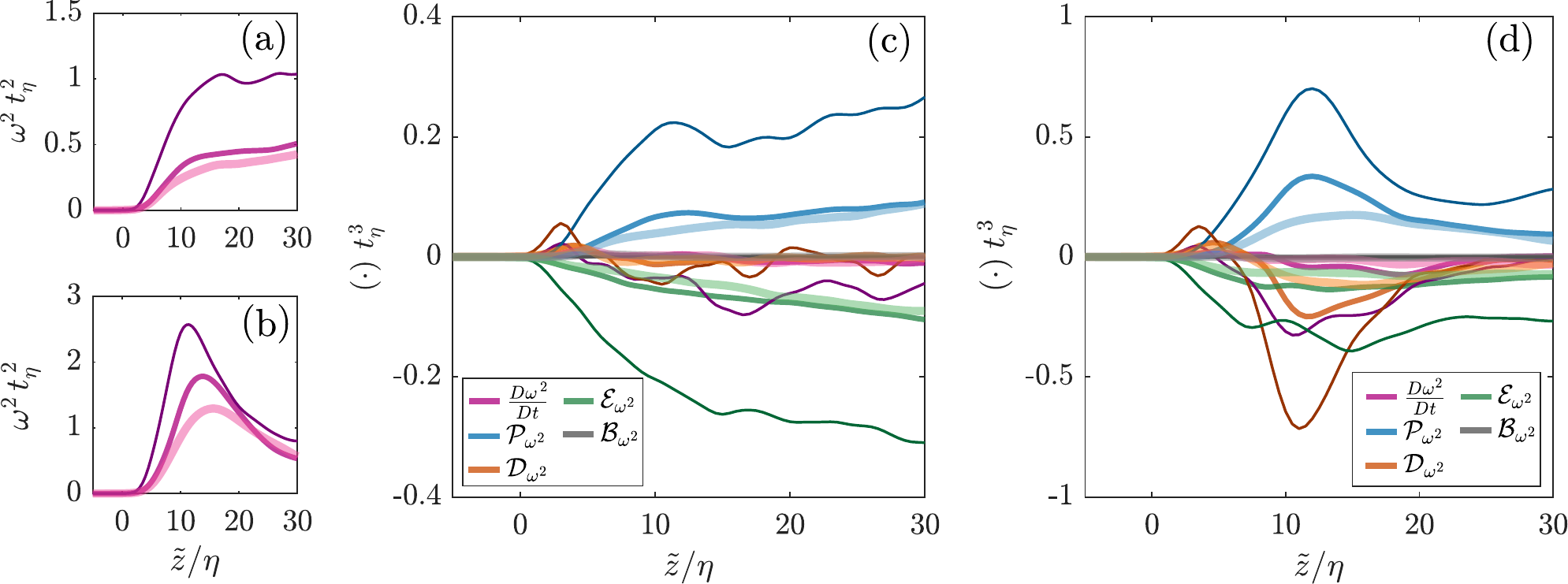}}
  \caption{Conditional average of the enstrophy (a) and of the terms in the enstrophy transport equation (b) with respect to the TNTI position. Conditional average of the enstrophy (c) and of the terms in the enstrophy transport equation (d) with respect to the presence of an OECS near the TNTI position. The thickness and the transparency of the lines increases with $Ri$.}
\label{fig:fig8}
\end{figure}

In figure \ref{fig:fig8} (a), we show the conditional average of the enstrophy with respect to the TNTI position.  The enstrophy has the typical profile described in the literature, with a sharp increase between $\tilde{z}/\eta=0$ and $\tilde{z}/\eta\approx 10$. As the stratification increases this jump is less sharp.

The conditional average profiles across the TNTI of the terms of the enstrophy transport equation are shown in figure \ref{fig:fig8}. The profiles present the typical shape described in the literature with the presence of three distinct regions. Near the TNTI, the vortex-stretching is negligible and the enstrophy increases due to viscous diffusion. This region, also known as the viscous superlayer (VSL), extends to $\tilde{z}/\eta\approx4$ where viscous diffusion is maximal. Further inside the turbulent region, the vortex stretching term increases and dominates the enstrophy growth, reaching a maximum at about $\tilde{z}/\eta\approx11$, where the viscous diffusion shows a minimum, while the dissipation term shows an inflection. The region between $\tilde{z}/\eta\approx4$ and $\tilde{z}/\eta\approx11$ is known as the turbulent sublayer (TSL). Form here onwards, the viscous diffusion becomes negligible and the vortex stretching is balanced by the viscous dissipation. This region is known as the turbulent core (TC). As also shown by \cite{krug2015turbulent}, at these $Ri$  the baroclinic torque is negligible near the TNTI.

In the recent literature \citep{watanabe2017role}, it has been inferred that the typical shape of the profiles of the enstrophy transport equation terms across the interface is compatible with the presence of vortical structures at about $\tilde{z}/\eta\approx9$. This distance was obtained by \citet{watanabe2017role}, as the sum between the radius $D/2\approx 5\eta$ of a Burger vortex modelling IVSs and the average size $\delta_{v}\approx 4\eta$ of the VSL. By superimposing the profiles of enstrophy transport equation terms conditioned with respect to the TNTI to those from the model, they observed a reasonable agreement. To test if the TNTI profiles are compatible with the presence of coherent structures, in figure \ref{fig:fig8} (d), we show the same conditional analysis presented above, this time conditioned with respect to the presence of an OECS. All the terms of the enstrophy transport equation exhibit the same trends shown in figure \ref{fig:fig8}(c) although with a different magnitude. Note the different vertical axis limits in figures \ref{fig:fig8}(c) and \ref{fig:fig8}(d). In particular, in the VSL and the TSL, all the terms are magnified up to one order of magnitude with respect to the levels when conditioned on the TNTI only. The shape of the profiles shown in figure \ref{fig:fig8} (d) is consistent with the radial profiles of a coherent structure positioned at a mean distance of $\tilde{z}/\eta\approx11$. This is further supported by figure \ref{fig:fig8}(b), in which we display the conditioned profile of the enstrophy with respect to the TNTI position and the presence of the OECSs. The enstrophy profiles have a peak at about $\tilde{z}/\eta\approx11$. As shown in \cite{watanabe2017role}, this position is close to the center of the most dominant vortical structures in the proximity of the TNTI. As the stratification increases, the peak position is slightly farther form the TNTI, which indicates that on average the OECSs are more distant form the TNTI. The results presented here clearly indicate that a large amount of the intensity of the enstrophy transport equation terms near the TNTI is attributable to the OECSs in the interface proximity.

\begin{figure}
  \centerline{\includegraphics[width=1\linewidth]{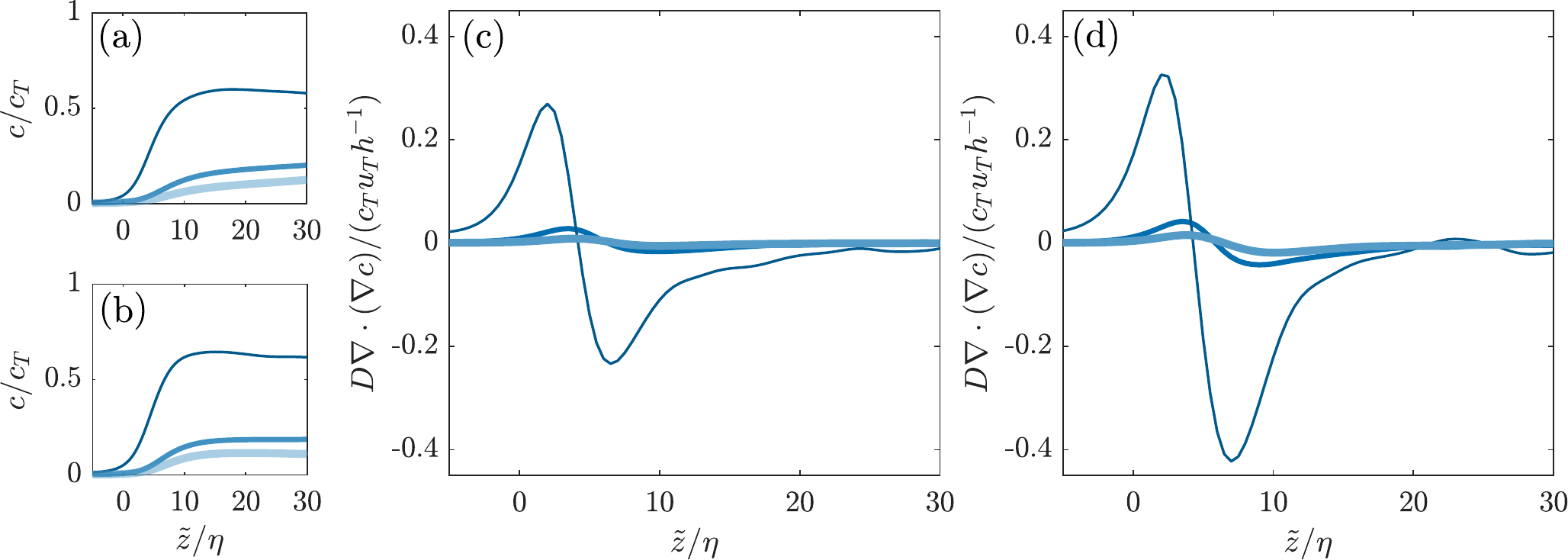}}
  \caption{Conditional average of the scalar (a) and of the terms in the scalar transport equation (c) with respect to the TNTI position. Conditional average of the scaler (b) and of the terms in the scalar transport equation (d) with respect to the presence of an OECS near the TNTI position.}
\label{fig:fig9}
\end{figure}

The same analysis conducted above is reproduced for the transport equation of the scalar. Figure \ref{fig:fig9}(a) shows that the scalar increases sharply from the non-turbulent to the turbulent side, which is steeper in the presence of OECSs (figure \ref{fig:fig9} b). That is, given that OECSs contain a higher scalar concentration as compared to the background, they also enhance the gradient of the scalar in the proximity of the TNTI. As the stratification increases, the concentration jump is less sharp. The conditional average of the scalar diffusion is shown in figure \ref{fig:fig9} (c). Notably, $D \nabla \cdot (\nabla c)$ is negative in the TC and positive in the TSL. It then changes sign and it shows a positive peak in the VSL. This means that $D \nabla \cdot (\nabla c)$ transports scalar from the TC to the outer fringes of the TNTI. Compatible with the curvature of the scalar profile, this is most effective for the unstratified case when compared to the gravity currents. By conditioning the profile of $D \nabla \cdot (\nabla c)$ with respect to the presence of an OECS, it can be seen that magnitude of the positive and negative peaks near the interface are approximately twice as high. This shows that the OECSs contribute to enhance the scalar diffusion near TNTI.    

\section{\label{sec:summary}Summary and conclusions}

We investigated the role of vortical structures on the transport of the enstrophy and of the scalar concentration. To educe vortical structures we employed an objective Eulerian coherent structures (OECSs) extraction method based on the instantaneous vorticity deviation field (IVD). 

We showed that high levels of enstrophy can be found within the boundaries of the OECSs. In particular, we showed that the radial profile of the enstrophy has a shape that is reminiscent of a half-Gaussian curve with a maximum at the center of the OECS followed by a sharp decrease towards the boundaries. For highly stable vortical structures, as in the case of the worms investigated by \cite{jimenez1993structure,da2011intense} the radial profile of enstrophy has been shown to be Gaussian. In their work, \citet{jimenez1993structure} identify these worms as regions of space that possess the highest vorticity in the flow (1\% of of the total volume). In our approach, even though the vortical structures are identified through a different extraction method that is independent of the vorticity magnitude, the results are very similar to those of \citet{jimenez1993structure}. 

The shape of the radial profile of the enstrophy is crucial to understand the time evolution of OECSs, in that it is at the base of the various terms of the enstrophy transport equation. The shape of the stretching and of the viscous diffusion terms are similar to those described in the literature for IVSs \cite{jimenez1993structure}, in which viscous diffusion of enstrophy is balanced by the vortex stretching. While in the previous literature these terms were only inferred based on a fitting to the Burgers model \citep{jimenez1993structure, da2011intense}, here these terms were computed directly from the flow fields. Moreover, contrary to the Burgers model, we showed that the viscous dissipation of enstrophy is not negligible. We also found that $D\omega^2/Dt$ is on average negative inside the OECSs. This can be explained in terms of vertical movement of the OECSs. By moving upwards, the structures carry enstrophy towards the outer layer of the current where on average enstrophy production cannot sustain the combined effect of viscous diffusion and destruction. As $Ri$ increases, the depth of the current is lower and the mean vorticity is higher. This is reflected on the OECSs enstrophy content which is higher for higher $Ri$. As a consequence, both the radial gradient and the radial curvature of enstrophy are higher and thus the magnitude of terms of the enstrophy transport equation is stronger. Importantly, we observed that for the gravity currents, the baroclinic torque is negligible in comparison to the other terms of the enstrophy equation.
 
By investigating the radial profile of the scalar, we showed that it resembles that of the enstrophy except that it is flatter. This means that, the OECSs on average carry a slightly higher scalar concentration compared with the background fluid. This can be explained in terms of material barriers. Since, on average, OECSs move upwards (figure \ref{fig:fig3}) and thus towards regions with a lower average concentration, a higher content of the scalar is expected within their boundaries as compared to the fluid around them. As the stratification increases, the radial profile of the scalar is less steep. This may, at first sight, be counter intuitive as $c$ is a passive scalar for $Ri0$, while it is an active one for $Ri11$ and $Ri22$. In the gravity currents, fluid parcels that possess a higher concentration are lighter and thus one might expect that these lighter parcels migrate towards the center of the OECSs due to centrifugal effects, which should result in a steeper radial profile as compared to $Ri0$. The apparent contradiction may be again explained by vertical movement of the OECSs. Since the structures move on average away from the wall faster in the unstratified case and traverse a steeper concentration gradient, higher differences between the concentration at the center of the OECSs and the surrounding fluid are seen for $Ri0$ as compared to $Ri11$ and $Ri22$.Consistent with the radial profiles of concentration (i.e. their curvature), we also observed that the diffusion of the concentration is positive inside the boundaries of the OECSs, that is, these structures contribute to redistribute the scalar across their boundaries.

Conditioning on the wall-normal position of OECSs center, we observed that the structures near the wall are the most active ones in terms of enstrophy transport. This is not unexpected given the highest gradient of the mean flow in this region. However, a difference we show is that the scalar diffusion is maximum in the outer region. This region corresponds to the maximum upward average velocity of the OECSs, which lends further support to the view that the scalar diffusion is connected to the vertical movement of the structures away from the wall.

The effect of vortical structures on the typical profiles of the enstrophy transport equation conditioned with respect to the TNTI position was investigated by further conditioning the analysis to the presence of OECSs. We showed that with this further condition, the intensity of the terms in the enstrophy transport equation can be one order of magnitude higher. This means that the shape of these profiles across the TNTI might be attributable to vortical structures near the interface. Recently, \citet{watanabe2017role} used a stable Burgers vortex positioned at approximately $9\eta$ form TNTI to reconstruct relatively well the vortex-stretching and the viscous diffusion profiles across the TNTI. However, in their model the viscous dissipation of enstrophy and  the effect of the baroclinic torque are missing. As a difference, with their model we used conditional analysis to assess the impact of the OECSs on all the terms of the enstrophy transport equation near the TNTI. We confirmed their findings related to the vortex-stretching and the viscous diffusion of enstrophy profiles, even though the profiles are compatible with an OECSs positioned a little farther away from the TNTI, i.e. at approximately $11\eta$. This difference may be related to the fact that their Burgers vortex models IVSs, while we employed OECSs that have a different size compared to IVSs. We also showed that while OECSs have no significant impact on the baroclinic torque which remains small in magnitude compared to the other terms, their footprint on the viscous dissipation is not negligible. This means that the model proposed by \citep{watanabe2017role} could be improved by further taking in account role of vortical structure on the viscous dissipation of enstrophy. Furthermore, compatible with the observations of previous literature \citep{krug2015turbulent,van2018mixing}, as the stratification increases, the magnitude of the enstrophy transport equation terms decreases, that is, the OECSs are less 'active' and entrainment process is less effective. 

We showed that similar considerations made for the enstrophy transport equation also hold for the scalar transport equation. When the profiles computed by conditioning to the TNTI position are further conditioned to the presence of OECSs, their magnitude is approximately double. Therefore, the OECSs also play an important role on the diffusion of the scalar across the TNTI by maintaining a steep concentration profile of the scalar.

The methodology used in this work to describe the enstrophy dynamics inside OECSs can be easily extended to other flows. It would be useful to understand if some characteristics of OECSs educed from the flows investigated here share common features with OECSs in other flows. We expect that for free shear flows such as wakes and jets or flows with weak unstable stratification such as turbulent convections in which the baroclinic torque has a weak influence on the enstrophy transport \citep{holzner2017turbulent}, the enstrophy dynamics of OECSs may be similar to that described here. When the baroclinic torque starts to acquire importance with respect to the other terms of the enstrophy transport equation, as in the case of turbulent plumes \citep{krug2017plumes}, the analysis introduced here can be replicated to understand how this term affects the enstrophy dynamics of OECSs and how it impacts enstrophy, scalar transport and the entrainment process near the TNTI.\\   

\textbf{Disclosure statement}

No potential conflict of interest was reported by the author(s).

\textbf{Acknowledgements}\\

We are grateful for financial support from DFG priority program SPP 1881 under grant number HA 7497/1-1.\\

J.-P.M. and M.v.R. were supported by the EPSRC project Multi-scale Dynamics at the Turbulent/Non-turbulent Interface of Jets and Plumes (grant number EP/R043175/1) and the UK Turbulence consortium (grant number EP/R029326/1).

\bibliographystyle{plainnat}

\begin{thebibliography}{53}
\providecommand{\natexlab}[1]{#1}
\providecommand{\url}[1]{\texttt{#1}}
\expandafter\ifx\csname urlstyle\endcsname\relax
  \providecommand{\doi}[1]{doi: #1}\else
  \providecommand{\doi}{doi: \begingroup \urlstyle{rm}\Url}\fi

\bibitem[Beta et~al.(2003)Beta, Schneider, and Farge]{beta2003wavelet}
C.~Beta, K.~Schneider, and M.~Farge.
\newblock Wavelet filtering to study mixing in 2d isotropic turbulence.
\newblock \emph{Comm. Nonlin. Sci. Num. Sim.}, 8\penalty0 (3-4):\penalty0
  537--545, 2003.

\bibitem[Bisset et~al.(2002)Bisset, Hunt, and Rogers]{bisset2002turbulent}
D.~K. Bisset, J.~C.~R. Hunt, and M.~M. Rogers.
\newblock The turbulent/non-turbulent interface bounding a far wake.
\newblock \emph{J.~Fluid Mech.}, 451:\penalty0 383--410, 2002.

\bibitem[Chauhan et~al.(2014)Chauhan, Philip, de~Silva, Hutchins, and
  Marusic]{chauhan2014turbulent}
K.~Chauhan, J.~Philip, C.~M. de~Silva, N.~Hutchins, and I.~Marusic.
\newblock The turbulent/non-turbulent interface and entrainment in a boundary
  layer.
\newblock \emph{J.~Fluid Mech.}, 742:\penalty0 119--151, 2014.

\bibitem[Corrsin and Kistler(1955)]{corrsin1955free}
S.~Corrsin and A.~L. Kistler.
\newblock Free-stream boundaries of turbulent flows.
\newblock 1955.

\bibitem[Craske and van Reeuwijk(2015)]{craske2015energy}
J.~Craske and M.~van Reeuwijk.
\newblock Energy dispersion in turbulent jets. part 1. direct simulation of
  steady and unsteady jets.
\newblock \emph{J.~Fluid Mech.}, 763:\penalty0 500--537, 2015.

\bibitem[da~Silva et~al.(2011)da~Silva, Dos~Reis, and Pereira]{da2011intense}
C.~B. da~Silva, R.J.N. Dos~Reis, and J.C.F. Pereira.
\newblock The intense vorticity structures near the turbulent/non-turbulent
  interface in a jet.
\newblock \emph{J.~Fluid Mech.}, 685:\penalty0 165--190, 2011.

\bibitem[da~Silva et~al.(2014)da~Silva, Hunt, Eames, and
  Westerweel]{da2014interfacial}
C.~B. da~Silva, J.~C.~R. Hunt, I.~Eames, and J.~Westerweel.
\newblock Interfacial layers between regions of different turbulence intensity.
\newblock \emph{Annu. Rev. Fluid Mech.}, 46:\penalty0 567--590, 2014.

\bibitem[Debusschere and Rutland(2004)]{debusschere2004turbulent}
B.~Debusschere and C.J. Rutland.
\newblock Turbulent scalar transport mechanisms in plane channel and couette
  flows.
\newblock \emph{Int. J. Heat Mass Transfer}, 47\penalty0 (8-9):\penalty0
  1771--1781, 2004.

\bibitem[Dharmarathne et~al.(2018)Dharmarathne, Pulletikurthi, and
  Castillo]{dharmarathne2018coherent}
S.~Dharmarathne, V.~Pulletikurthi, and L.~Castillo.
\newblock Coherent vortical structures and their relation to hot/cold spots in
  a thermal turbulent channel flow.
\newblock \emph{Fluid}, 3\penalty0 (1):\penalty0 14, 2018.

\bibitem[Dimotakis(2000)]{dimotakis2000mixing}
P.~E. Dimotakis.
\newblock The mixing transition in turbulent flows.
\newblock \emph{J.~Fluid Mech.}, 409:\penalty0 69--98, 2000.

\bibitem[Dubief and Delcayre(2000)]{dubief2000coherent}
Y.~Dubief and F.~Delcayre.
\newblock On coherent-vortex identification in turbulence.
\newblock \emph{J. Turb.}, 1\penalty0 (1):\penalty0 011--011, 2000.

\bibitem[Ellison and Turner(1959)]{ellison1959turbulent}
T.~H. Ellison and J.~S. Turner.
\newblock Turbulent entrainment in stratified flows.
\newblock \emph{J.~Fluid Mech.}, 6\penalty0 (3):\penalty0 423--448, 1959.

\bibitem[Ellison(1957)]{ellison1957turbulent}
T.H. Ellison.
\newblock Turbulent transport of heat and momentum from an infinite rough
  plane.
\newblock \emph{J.~Fluid Mech.}, 2\penalty0 (5):\penalty0 456--466, 1957.

\bibitem[Frisch and Kolmogorov(1995)]{frisch1995turbulence}
U.~Frisch and A.~N. Kolmogorov.
\newblock \emph{Turbulence: the legacy of AN Kolmogorov}.
\newblock Cambridge university press, 1995.

\bibitem[Fr{\"o}hlich et~al.(2008)Fr{\"o}hlich, Garc{\'\i}a-Villalba, and
  Rodi]{frohlich2008scalar}
J.~Fr{\"o}hlich, M.~Garc{\'\i}a-Villalba, and W.~Rodi.
\newblock Scalar mixing and large-scale coherent structures in a turbulent
  swirling jet.
\newblock \emph{Flow Turbul. Combust.}, 80\penalty0 (1):\penalty0 47--59, 2008.

\bibitem[Ganapathisubramani et~al.(2008)Ganapathisubramani, Lakshminarasimhan,
  and Clemens]{ganapathisubramani2008investigation}
B.~Ganapathisubramani, K.~Lakshminarasimhan, and N.~T. Clemens.
\newblock Investigation of three-dimensional structure of fine scales in a
  turbulent jet by using cinematographic stereoscopic particle image
  velocimetry.
\newblock \emph{J.~Fluid Mech.}, 598:\penalty0 141--175, 2008.

\bibitem[Haller(2015)]{haller2015lagrangian}
G.~Haller.
\newblock Lagrangian coherent structures.
\newblock \emph{Annu. Rev. Fluid Mech.}, 47:\penalty0 137--162, 2015.

\bibitem[Haller(2016)]{haller2016dynamic}
G.~Haller.
\newblock Dynamic rotation and stretch tensors from a dynamic polar
  decomposition.
\newblock \emph{J. Mech. Phys. Solids}, 86:\penalty0 70--93, 2016.

\bibitem[Haller et~al.(2016)Haller, Hadjighasem, Farazmand, and
  Huhn]{haller2016defining}
G.~Haller, A.~Hadjighasem, M.~Farazmand, and F.~Huhn.
\newblock Defining coherent vortices objectively from the vorticity.
\newblock \emph{J.~Fluid Mech.}, 795:\penalty0 136--173, 2016.

\bibitem[Holzner and L{\"u}thi(2011)]{holzner2011laminar}
M.~Holzner and B.~L{\"u}thi.
\newblock Laminar superlayer at the turbulence boundary.
\newblock \emph{Phys. Rev. Lett.}, 106\penalty0 (13):\penalty0 134503, 2011.

\bibitem[Holzner and van Reeuwijk(2017)]{holzner2017turbulent}
M.~Holzner and M.~van Reeuwijk.
\newblock The turbulent/nonturbulent interface in penetrative convection.
\newblock \emph{J. Turbul.}, 18\penalty0 (3):\penalty0 260--270, 2017.

\bibitem[Holzner et~al.(2007)Holzner, Liberzon, Nikitin, Kinzelbach, and
  Tsinober]{holzner2007small}
M.~Holzner, A.~Liberzon, N.~Nikitin, W.~Kinzelbach, and A.~Tsinober.
\newblock Small-scale aspects of flows in proximity of the
  turbulent/nonturbulent interface.
\newblock \emph{Phys. Fluids}, 19\penalty0 (7):\penalty0 071702, 2007.

\bibitem[Holzner et~al.(2008)Holzner, Liberzon, Nikitin, L{\"u}thi, Kinzelbach,
  and Tsinober]{holzner2008lagrangian}
M.~Holzner, A.~Liberzon, N.~Nikitin, B.~L{\"u}thi, W.~Kinzelbach, and
  A.~Tsinober.
\newblock A lagrangian investigation of the small-scale features of turbulent
  entrainment through particle tracking and direct numerical simulation.
\newblock \emph{J.~Fluid Mech.}, 598:\penalty0 465--475, 2008.

\bibitem[Hua and Klein(1998)]{hua1998exact}
B.L. Hua and P.~Klein.
\newblock An exact criterion for the stirring properties of nearly
  two-dimensional turbulence.
\newblock \emph{Physica D}, 113\penalty0 (1):\penalty0 98--110, 1998.

\bibitem[Hunt et~al.(1988)Hunt, Wray, and Moin]{hunt1988eddies}
J.~C.~R. Hunt, A.~A. Wray, and P.~Moin.
\newblock Eddies, streams, and convergence zones in turbulent flows.
\newblock 1988.

\bibitem[Hussain(1986)]{hussain1986coherent}
A.~K. M.~F. Hussain.
\newblock Coherent structures and turbulence.
\newblock \emph{J.~Fluid Mech.}, 173:\penalty0 303--356, 1986.

\bibitem[Jimenez and Wray(1998)]{jimenez1998characteristics}
J.~Jimenez and A.~A. Wray.
\newblock On the characteristics of vortex filaments in isotropic turbulence.
\newblock \emph{J.~Fluid Mech.}, 373:\penalty0 255--285, 1998.

\bibitem[Jim{\'e}nez et~al.(1993)Jim{\'e}nez, Wray, Saffman, and
  Rogallo]{jimenez1993structure}
J.~Jim{\'e}nez, A.~A Wray, P.~G. Saffman, and R.~S. Rogallo.
\newblock The structure of intense vorticity in isotropic turbulence.
\newblock \emph{J.~Fluid Mech.}, 255:\penalty0 65--90, 1993.

\bibitem[Kadoch et~al.(2011)Kadoch, Iyer, Donzis, Schneider, Farge, and
  Yeung]{kadoch2011role}
B.~Kadoch, K.~Iyer, D.~Donzis, K.~Schneider, M.~Farge, and P.K. Yeung.
\newblock On the role of vortical structures for turbulent mixing using direct
  numerical simulation and wavelet-based coherent vorticity extraction.
\newblock \emph{J.Turbul.}, \penalty0 (12):\penalty0 N20, 2011.

\bibitem[Kang et~al.(2007)Kang, Tanahashi, and Miyauchi]{kang2007dynamics}
S.J. Kang, M.~Tanahashi, and T.~Miyauchi.
\newblock Dynamics of fine scale eddy clusters in turbulent channel flows.
\newblock \emph{J. Turbul.}, \penalty0 (8):\penalty0 N52, 2007.

\bibitem[Krug et~al.(2015)Krug, Holzner, L{\"u}thi, Wolf, Kinzelbach, and
  Tsinober]{krug2015turbulent}
D.~Krug, M.~Holzner, B.~L{\"u}thi, M.~Wolf, W.~Kinzelbach, and A.~Tsinober.
\newblock The turbulent/non-turbulent interface in an inclined dense gravity
  current.
\newblock \emph{J.~Fluid Mech.}, 765:\penalty0 303--324, 2015.

\bibitem[Krug et~al.(2017{\natexlab{a}})Krug, Chung, Philip, and
  Marusic]{krug2017plumes}
D.~Krug, D.~Chung, J.~Philip, and I.~Marusic.
\newblock Global and local aspects of entrainment in temporal plumes.
\newblock \emph{J.~Fluid Mech.}, 812:\penalty0 222--250, 2017{\natexlab{a}}.

\bibitem[Krug et~al.(2017{\natexlab{b}})Krug, Holzner, Marusic, and van
  Reeuwijk]{krug2017fractal}
D.~Krug, M.~Holzner, I.~Marusic, and M.~van Reeuwijk.
\newblock Fractal scaling of the turbulence interface in gravity currents.
\newblock \emph{J.~Fluid Mech.}, 820, 2017{\natexlab{b}}.

\bibitem[Lesieur(1987)]{lesieur1987turbulence}
Ml. Lesieur.
\newblock \emph{Turbulence in fluids: stochastic and numerical modelling}.
\newblock Nijhoff Boston, MA, 1987.

\bibitem[Mistry et~al.(2016)Mistry, Philip, Dawson, and
  Marusic]{mistry2016entrainment}
D.~Mistry, J.~Philip, J.~R. Dawson, and I.~Marusic.
\newblock Entrainment at multi-scales across the turbulent/non-turbulent
  interface in an axisymmetric jet.
\newblock \emph{J.~Fluid Mech.}, 802:\penalty0 690--725, 2016.

\bibitem[Neamtu-Halic et~al.(2019)Neamtu-Halic, Krug, Haller, and
  Holzner]{neamtu2019lagrangian}
M.~M. Neamtu-Halic, D.~Krug, G.~Haller, and M.~Holzner.
\newblock Lagrangian coherent structures and entrainment near the
  turbulent/non-turbulent interface of a gravity current.
\newblock \emph{J.~Fluid Mech.}, 877:\penalty0 824--843, 2019.

\bibitem[Neamtu-Halic et~al.(2020)Neamtu-Halic, Krug, Mollicone, van Reeuwijk,
  Haller, and Holzner]{neamtu2020evolution}
M.~M. Neamtu-Halic, D.~Krug, J.-P. Mollicone, M.~van Reeuwijk, G.~Haller, and
  M.~Holzner.
\newblock Evolution of the turbulence interface in flows with and without
  stable stratification.
\newblock \emph{arXiv preprint arXiv:2001.02427}, 2020.

\bibitem[Okubo(1970)]{okubo1970horizontal}
A.~Okubo.
\newblock Horizontal dispersion of floatable particles in the vicinity of
  velocity singularities such as convergences.
\newblock In \emph{Deep-Sea Res.}, volume~17, pages 445--454. Elsevier, 1970.

\bibitem[Serra and Haller(2016)]{serra2016objective}
Mattia Serra and George Haller.
\newblock Objective eulerian coherent structures.
\newblock \emph{Chaos}, 26\penalty0 (5):\penalty0 053110, 2016.

\bibitem[Siggia(1981)]{siggia1981numerical}
E.~D. Siggia.
\newblock Numerical study of small-scale intermittency in three-dimensional
  turbulence.
\newblock \emph{J.~Fluid Mech}, 107:\penalty0 375--406, 1981.

\bibitem[Silva et~al.(2018)Silva, Zecchetto, and da~Silva]{silva2018scaling}
T.~S. Silva, M.~Zecchetto, and C.~B. da~Silva.
\newblock The scaling of the turbulent/non-turbulent interface at high reynolds
  numbers.
\newblock \emph{J.~Fluid Mech.}, 843:\penalty0 156--179, 2018.

\bibitem[Tanahashi et~al.(2001)Tanahashi, Iwase, and
  Miyauchi]{tanahashi2001appearance}
M.~Tanahashi, S.~Iwase, and T.~Miyauchi.
\newblock Appearance and alignment with strain rate of coherent fine scale
  eddies in turbulent mixing layer.
\newblock \emph{J. Turbul.}, 2\penalty0 (6):\penalty0 1--17, 2001.

\bibitem[Taveira et~al.(2013)Taveira, D., Lopes, and
  da~Silva]{taveira2013lagrangian}
R.~R. Taveira, J.~S. D., D.~C. Lopes, and C.~B. da~Silva.
\newblock Lagrangian statistics across the turbulent-nonturbulent interface in
  a turbulent plane jet.
\newblock \emph{Phys. Rev. E}, 88\penalty0 (4):\penalty0 043001, 2013.

\bibitem[Townsend(1958)]{townsend1958turbulent}
A.A. Townsend.
\newblock Turbulent flow in a stably stratified atmosphere.
\newblock \emph{J.~Fluid Mech.}, 3\penalty0 (4):\penalty0 361--372, 1958.

\bibitem[Tsinober(2009)]{tsinober2009informal}
A.~Tsinober.
\newblock \emph{An informal conceptual introduction to turbulence}, volume 483.
\newblock Springer, 2009.

\bibitem[van Reeuwijk et~al.(2018)van Reeuwijk, Krug, and
  Holzner]{van2018small}
M.~van Reeuwijk, D.~Krug, and M.~Holzner.
\newblock Small-scale entrainment in inclined gravity currents.
\newblock \emph{Environ. Fluid Mech.}, 18\penalty0 (1):\penalty0 225--239,
  2018.

\bibitem[van Reeuwijk et~al.(2019)van Reeuwijk, Holzner, and
  Caulfield]{van2018mixing}
M.~van Reeuwijk, M.~Holzner, and C.~P. Caulfield.
\newblock Mixing and entrainment are suppressed in inclined gravity currents.
\newblock \emph{J.~Fluid Mech.}, 873:\penalty0 786--815, 2019.

\bibitem[Vincent and Meneguzzi(1991)]{vincent1991satial}
A.~Vincent and M.~Meneguzzi.
\newblock The satial structure and statistical properties of homogeneous
  turbulence.
\newblock \emph{J.~Fluid Mech.}, 225:\penalty0 1--20, 1991.

\bibitem[Watanabe et~al.(2015)Watanabe, Sakai, Nagata, Ito, and
  Hayase]{watanabe2015turbulent}
T.~Watanabe, Y.~Sakai, K.~Nagata, Y.~Ito, and T.~Hayase.
\newblock Turbulent mixing of passive scalar near turbulent and non-turbulent
  interface in mixing layers.
\newblock \emph{Phys. Fluid}, 27\penalty0 (8):\penalty0 085109, 2015.

\bibitem[Watanabe et~al.(2016)Watanabe, da~Silva, Sakai, Nagata, and
  Hayase]{watanabe2016lagrangian}
T.~Watanabe, C.~B. da~Silva, Y.~Sakai, K.~Nagata, and T.~Hayase.
\newblock Lagrangian properties of the entrainment across
  turbulent/non-turbulent interface layers.
\newblock \emph{Phys. Fluid.}, 28\penalty0 (3):\penalty0 031701, 2016.

\bibitem[Watanabe et~al.(2017)Watanabe, Jaulino, Taveira, da~Silva, Nagata, and
  Sakai]{watanabe2017role}
T.~Watanabe, R.~Jaulino, R.R. Taveira, C.B. da~Silva, K.~Nagata, and Y.~Sakai.
\newblock Role of an isolated eddy near the turbulent/non-turbulent interface
  layer.
\newblock \emph{Phy. Rev. Fluid}, 2\penalty0 (9):\penalty0 094607, 2017.

\bibitem[Weiss(1991)]{weiss1991dynamics}
J.~Weiss.
\newblock The dynamics of enstrophy transfer in two-dimensional hydrodynamics.
\newblock \emph{Physica D}, 48\penalty0 (2-3):\penalty0 273--294, 1991.

\bibitem[Westerweel et~al.(2009)Westerweel, Fukushima, Pedersen, and
  Hunt]{westerweel2009momentum}
J.~Westerweel, C.~Fukushima, J.~M. Pedersen, and J.~C.~R. Hunt.
\newblock Momentum and scalar transport at the turbulent/non-turbulent
  interface of a jet.
\newblock \emph{J.~Fluid Mech.}, 631:\penalty0 199--230, 2009.

\end{thebibliography}

\end{document}